# Why do non-gauge invariant terms appear in the vacuum polarization tensor?

Author: Dan Solomon.
Email: dsolom2@uic.edu

**Abstract.**
It is well known that quantum field theory at the formal level is gauge invariant. However a calculation of the vacuum polarization tensor will include non-gauge invariant terms. These terms must be removed from the calculation in order to get a physically correct result. One common way to do this today is the technique of "dimensional regularization". It has recently been noted [2] that at one time a supersymmetric-like solution to the problem was explored – that is, the right combination of fields would cause the offending terms to cancel out. I will examine some of this early work and pose the question – why do the non-gauge invariant terms appear in the first place? I will show that this is due to an improper mathematical step in the formulation of the perturbative expansion. I will then show that when this step is corrected the result is gauge invariant. However a supersymmetric-like solution is still required to cancel out a divergent term.

## 1. Introduction.

Supersymmetry, as currently conceived, can trace its origins back to the 1970s [1]. However a recent paper by C. Jarlskog [2] reveals that a supersymmetric-like solution to a problem related to the calculation of the vacuum polarization tensor was suggested as far back as the late 1940's by various researchers (Refer to [2] for references). The problem in question was the fact that calculations of the vacuum polarization tensor produced a term that was both divergent and non-gauge invariant. This term was obviously non-physical and had to be removed from the solutions in order to obtain a physically correct theory. It was shown that this term would cancel for the right combination of fields with the correct masses.

One thing that is disturbing about this result is that fact that these calculations give non-gauge invariant results. This is despite the fact that quantum field theory is supposed to be gauge invariant. I will review some of this work and show that the non-gauge invariant term is a result of an improper mathematical step and that when this is corrected all terms are gauge invariant. However there will still be a divergent term which can be eliminated by the correct combination of fields so that a supersymmetric solution is still required.

In the following discussion $\hbar = c = 1$. Also I will use a somewhat "old fashioned" notational convention of Ref. [3] where $ab = a_1b_1 + a_2b_2 + a_3b_3 - a_0b_0$ and $x_4 = ix_0$ where $x_0 = t$. This convention is used because it is consistent with most of the references cited herein. Also in this convention the position of the indices does not matter. That is $a_\mu$ is the same as $a^\mu$. For consistency I have made all indices as subscripts.

## 2. The problem of vacuum polarization.

The electromagnetic field tensor is $F_{uv} = (\partial_\mu A_v - \partial_v A_\mu)$ where $A_v$ is the electric potential. A change in the gauge is a change in electric potential which does not produce a change in $F_{uv}$. Such a change is given by $A_v \to A_v + \partial_v \chi$ where $\chi(x)$ is an arbitrary function.

In order for quantum field theory to be gauge invariant a change in the gauge cannot produce a change in any physical observable such as the current and charge expectation values. For example the first order change in the vacuum current, due to an applied electromagnetic field, can be shown to be,

$$j^{(1)}_{\mu,vac}(x) = \int \pi_{\mu v}(x-x') A_v(x') d^4x' \tag{2.1}$$

where $\pi_{\mu v}$ is the vacuum polarization tensor. We can write this equation in momentum space as,

$$j^{(1)}_{\mu,vac}(p) = \pi_{\mu v}(p) A_v(p) \tag{2.2}$$

In this case the gauge transformation takes the form,

$$A_v(p) \to A_v(p) + ip_v \chi(p) \tag{2.3}$$

Using the above the change in the vacuum current due to the gauge transformation is,

$$\delta_g j^{(1)}_{\mu,vac}(p) = ip_v \pi_{\mu v}(p) \chi(p) \tag{2.4}$$

For QFT to be gauge invariant $\delta_g j^{(1)}_{\mu,vac}(p)$ must be zero. Therefore the vacuum polarization tensor must satisfy the relationship,

$$p_v \pi_{\mu v}(p) = 0 \tag{2.5}$$

Despite this is has been well established that when the vacuum polarization tensor is calculated non-gauge invariant terms appear in the result so that (2.5) does not hold. (See discussion in D. Solomon [4]). For example a calculation of the vacuum polarization tensor by W. Heitler (see p322 of [5]) yields,

$$\pi_{\mu v}(p) = \pi^G_{\mu v}(p) + \pi^{NG}_{\mu v}(p) \tag{2.6}$$

where,

$$\pi^G_{\mu v}(p) = \left(\frac{1}{4\pi}\right)\left(\frac{2e^2}{3\pi}\right)(p_u p_v - \delta_{\mu v} p^2) \times \int_{2m}^{\infty} dz \frac{(z^2 + 2m^2)\sqrt{(z^2 - 4m^2)}}{z^2(z^2 + p^2)} \tag{2.7}$$

$$\pi^{NG}_{\mu v}(p) = \left(\frac{1}{4\pi}\right)\left(\frac{2e^2}{3\pi}\right)\delta_{\mu v}(1-\delta_{\mu 4}) \times \int_{2m}^{\infty} dz \frac{(z^2 + 2m^2)\sqrt{(z^2 - 4m^2)}}{z^2} \tag{2.8}$$

The term $\pi^G_{\mu v}(p)$ is gauge invariant because $p_\mu \pi^G_{\mu v}(p) = 0$. However the term $\pi^{NG}_{\mu v}(p)$ is not gauge invariant because $p_\mu \pi^{NG}_{\mu v}(p) \neq 0$. In order to obtain a physically valid result it is necessary to "correct" equation (2.6) by removing the non-gauge invariant term $\pi^{NG}_{\mu v}(p)$ from the result. (Note – the $(1/4\pi)$ term does not appear in [5]. I have added this because Heitler works in Gaussian units and the work here will be in natural units).

Another example is given in section 6.4 of Nishijima [3]. Here it is shown that the vacuum polarization includes a non-gauge invariant term which must be removed to obtain the "correct" gauge invariant result. According to Nishijima (pages 218-220 of [3]) the term that is discarded is $\pi_{\mu\nu}^{NG}(p) = \delta_{\mu\nu}(A + Bp^2)$ where $A$ is a quadratically divergent constant and $B$ is a finite constant.

Sakurai (Section 4-7 of Ref [6]) argues that based on requirements of Lorentz covariance the vacuum polarization tensor should be of the form,

$$\pi^{\mu\nu}(p) = D\delta_{\mu\nu} + \delta_{\mu\nu}p^2\pi^{(1)}(p^2) + p_\mu p_\nu \pi^{(2)}(p^2) \qquad (2.9)$$

Here $D$ is a constant which would correspond to a photon mass which means that $D$ must be zero. Sakura then shows that $D$ is given by,

$$D = ie^2 \int \frac{d^4p}{(2\pi)^4} \frac{(2p^2 + 4m^2)}{(p^2 + m^2 - i\varepsilon)^2} \qquad (2.10)$$

Sakurai writes that "It is not difficult to convince oneself that almost any 'honest' calculation gives $D \neq 0$."

Another example is from Section 14.2 of Greiner et al [7] who obtain the following expression for the vacuum polarization tensor,

$$\pi^{\mu\nu}(p) = (g^{\mu\nu}p^2 - p^\mu p^\nu)\pi(p^2) + g^{\mu\nu}\pi_{sp}(p^2) \qquad (2.11)$$

where $\pi(p^2)$ and $\pi_{sp}(p^2)$ are given in [7]. It is evident that the above expression is not gauge invariant unless $\pi_{sp}(p^2)$ is zero. In [7] it is shown that $\pi_{sp}(p^2) \neq 0$. Therefore, in order to obtain the physically "correct" gauge invariant result this term must be dropped.

This is just a small sample of a vast literature on the subject. Virtually all derivations of the vacuum current show the existence of a non-gauge invariant term. Of course in order to get a result that ultimately agrees with experiment these non-gauge invariant terms must be removed. Different authors use different approaches. As mentioned above an early approach used a supersymmetric-like solution. Jost and Rayski [8] showed that the non-gauge invariant terms from Dirac and scalar fields tend to cancel out. The combination of fields that will give exact cancelation is,

$$N = 2n; \quad \sum_{i=1}^{N} M_i^2 = 2\sum_{i=1}^{n} m_i^2 \qquad (2.12)$$

where $M_i$ is the mass of the scalar field and $m_i$ is the mass of the Dirac field. Sakata and Umezawa [9] extend this formula to include vector fields. This approach was eventually abandoned (and forgotten) because, at the time, it was thought that there was no reason to believe that such a relationship existed. The history of this approach is discussed in more depth by Jarlsog [2].

Another possibility is to simply drop the offending term from the final result. Heitler[5], Nishijima[3] and Greiner [7] take this approach. The other approach is to develop mathematical methods to remove these terms. One such technique is "dimensional regularization" and is widely used today. An example of this is given by [10].

The obvious question to ask is the following - Why do non-gauge invariant terms appear in a theory that is supposed to be gauge invariant? Why don't we obtain a result that is manifestly gauge invariant? By manifestly gauge invariant I mean that the gauge invariance of some derived quantity such as the vacuum polarization tensor is obvious. Why is an additional step to needed to remove the non-gauge invariant term?

In the rest of this paper I will examine this problem and try to demonstrate what must be done at the fundamental level to produce a theory that is manifestly gauge invariant. My philosophical point of view is as follows – A scientific theory must be a correct model of the real world in that it must agree with experiment, but it also must be mathematically consistent. Quantum theory claims to be gauge invariant. However this property of gauge invariance is not born out when the vacuum polarization tensor is calculated. To make the theory gauge invariant an extra step is required to remove the offending terms. This extra step may be a "regularization" method such as dimensional regularization or, as is often the case, the offending non-gauge invariant terms are simply removed "by hand", that is, just dropped from the result. The result is that the theory, in the end, does agree with experiment. However my concern is that the theory is not mathematically consistent. A theory that claims to be gauge invariant should produce manifestly gauge invariant results and not produce non-gauge invariant terms that have to be removed at the end of the calculation by an additional step.

I will start by examining a relatively straightforward calculation of the vacuum polarization tensor by W. Pauli [11]. Here an equation for the first order change in the vacuum current of a Dirac field due to an external applied electromagnetic potential is obtained. Pauli shows that this equation is not gauge invariant. I will look in detail at Pauli's calculation and how show how to modify the field operator in order to achieve a gauge invariant result and then develop an expression for the vacuum current, $j_{D\mu}^{(1)}$, for the Dirac field which is manifestly gauge invariant. I will then calculate the vacuum current in momentum space for the simple case where the four momentum, $p_\mu$, of the applied electric potential satisfies $\vec{p}=0$ and $2m>|p_0|>0$ where $m$ is the mass of the Dirac field. The result is that $j_{D\mu}^{(1)}$ consists of a sum of three terms. The first is a finite term that is consistent with the standard result. The second is a logarithmically divergent term that can be identified as a charge renormalization term. This is also consistent with the standard result. The third term is a highly divergent term which is obviously non-physical. Even though it is non-physical it is mathematical correct. That is, it follows from the mathematical calculation. It is shown that this term can be eliminated through supersymmetry. If we assume that there are two charged scalar fields with identical mass to the Dirac field then it is shown that this divergent term cancels out when the contribution to the vacuum current by the scalar fields is included.

## 3. Vacuum Polarization from Pauli's lectures.

In this section we will examine a calculation of vacuum polarization as discussed by W. Pauli [11]. In Section 8 of [11] the vacuum polarization tensor for a Dirac field is calculated in the presence of an external field. The calculation is done in the Heisenberg picture so that the time dependence is associated with the field operator and the state vector is time independent. Referring to Section 8 of [11] the Dirac equation in the presence of a classical electromagnetic potential is,

$$\left(\gamma_\nu \frac{\partial}{\partial x_\nu} + m\right)\hat{\psi}(x) = ie\gamma_\nu A_\nu(x)\hat{\psi}(x) \tag{3.1}$$

where $\hat{\psi}(x)$ is the field operator. The free field solution is the solution where $A_\nu(x) = 0$. It is given in Section 12 of Ref [12] as,

$$\hat{\psi}_\alpha^{(0)}(x) = \frac{1}{\sqrt{V}} \sum_{\vec{q}} \left[ e^{i(\vec{q}\cdot\vec{x}-Et)} \sum_{r=1}^{2} \hat{a}_r(\vec{q}) u_\alpha^{(r)}(\vec{q}) + e^{i(\vec{q}\cdot\vec{x}+Et)} \sum_{r=3}^{4} \hat{a}_r(\vec{q}) u_\alpha^{(r)}(\vec{q}) \right] \tag{3.2}$$

where the $\hat{a}_r(\vec{q})$ obey the usual anticommutation relationships.

The current operator is given by,

$$\hat{j}_\mu(x) = -\frac{ie}{2}\left[\hat{\psi}_\alpha(x), \hat{\bar{\psi}}_\beta(x)\right]\gamma_{\mu,\alpha\beta} \tag{3.3}$$

In general we cannot solve (3.1) for an arbitrary $A_\nu(x)$. However, if $A_\nu(x) = \partial_\nu \chi(x)$ then a solution is possible. In this case the solution to the Dirac equation is,

$$\hat{\psi}(x) = e^{ie\chi(x)}\hat{\psi}^{(0)}(x); \quad \hat{\bar{\psi}}(x) = e^{-ie\chi(x)}\hat{\bar{\psi}}^{(0)}(x) \tag{3.4}$$

If $A_\nu(x) = \partial_\nu \chi(x)$ the electromagnetic field tensor $F_{\mu\nu} = 0$ so that $A_\nu(x) = \partial_\nu \chi(x)$ is simply a gauge transformation from zero electromagnetic field. When (3.4) is used in (3.3) we obtain,

$\hat{j}_\mu(x) = (-ie/2)\left[\hat{\psi}_\alpha^{(0)}(x), \hat{\bar{\psi}}_\beta^{(0)}(x)\right]\gamma_{\mu,\alpha\beta}$. Note that $\chi(x)$ does not appear in this expression.

Therefore a change in the gauge did not produce a change in the current operator which is a requirement of a gauge invariant theory.

So far it has been shown that Dirac field theory is gauge invariant. So why is there a problem with gauge invariance in perturbation theory? To start an examination of perturbation theory first consider an expansion of the exponential in (3.4) and to obtain,

$$\hat{\psi}(x) = (1 + ie\chi(x) + \ldots)\hat{\psi}^{(0)}(x); \quad \hat{\bar{\psi}}(x) = \hat{\bar{\psi}}^{(0)}(x)(1 - ie\chi(x) + \ldots) \tag{3.5}$$

Use this in (3.3) and retain the lowest order term to obtain,

$$\begin{aligned}\hat{j}_\mu(x) &= -\frac{ie}{2}\left[\hat{\psi}_\alpha^{(0)}(x)(1 + ie\chi(x) + \ldots), (1 - ie\chi(x) + \ldots)\hat{\bar{\psi}}_\beta^{(0)}(x)\right]\gamma_{\mu,\alpha\beta} \\ &= -\frac{ie}{2}\left[\hat{\psi}_\alpha^{(0)}(x), \hat{\bar{\psi}}_\beta^{(0)}(x)\right]\gamma_{\mu,\alpha\beta} + O(e^2)\end{aligned} \tag{3.6}$$

This result shows that the first order expansion of the theory is gauge invariant. This is, of course, what we would expect from a gauge invariant theory.

However the situation changes when we consider the first order expansion for arbitrary $A_\nu(x)$. In this case an exact solution is normally not possible and we have to use the perturbation theory. From Section 8 of [11] the expansion of the field operator is,

$$\hat{\psi}(x) = \hat{\psi}^{(0)}(x) + e\hat{\psi}^{(1)}(x) + \ldots \tag{3.7}$$

where $\hat{\psi}^{(1)}(x)$ is given by,

$$\left(\gamma_\nu \frac{\partial}{\partial x_\nu} + m\right) e\hat{\psi}^{(1)}(x) = ie\gamma_\nu A_\nu(x)\hat{\psi}^{(0)}(x) \tag{3.8}$$

When (3.7) is substituted into (3.3) we obtain that the first order change in the current operator is,

$$\hat{j}_\mu^{(1)}(x) = -\frac{ie^2}{2}\left(\left[\hat{\psi}_\alpha^{(1)}(x), \hat{\bar{\psi}}_\beta^{(0)}(x)\right] + \left[\hat{\psi}_\alpha^{(0)}(x), \hat{\bar{\psi}}_\beta^{(1)}(x)\right]\right)\gamma_{\mu,\alpha\beta} \tag{3.9}$$

The first order change in the vacuum current is then,

$$j_{D\mu}^{(1)}(x) = \langle 0|\hat{j}_\mu^{(1)}(x)|0\rangle \tag{3.10}$$

where $|0\rangle$ is the vacuum state and the subscript $D$ on $j_{D\mu}^{(1)}(x)$ indicates that this is the vacuum current for a Dirac field.

From Pauli [11] the solution to (3.8) is,

$$e\hat{\psi}^{(1)}(x) = -ie\int S^{ret}(x-x')\gamma_\nu A_\nu(x')\hat{\psi}^{(0)}(x')d^4x' \tag{3.11}$$

$$e\hat{\bar{\psi}}^{(1)}(x) = -ie\int \hat{\bar{\psi}}^{(0)}(x')\gamma_\nu A_\nu(x')S^{adv}(x'-x)d^4x' \tag{3.12}$$

where $S^{ret}(x)$ and $S^{adv}(x)$ are the retarded and advanced Greens function, respectively and are given by,

$$S^{ret}(x) = -\theta(t)S(x), \ S^{adv}(x) = \theta(-t)S(x) \tag{3.13}$$

where,

$$S(x) = \left(\gamma\frac{\partial}{\partial x} - m\right)\Delta(x), \ \Delta(\vec{x},t) = -\left(\frac{1}{2\pi}\right)^3 \int \exp[i\vec{k}\cdot\vec{x}]\frac{\sin\omega t}{\omega}d^3k, \ \omega = \sqrt{k^2+m^2} \tag{3.14}$$

When (3.11) and (3.12) are used in (3.9) and (3.10) the first order change in the vacuum current is,

$$j_{D\mu}^{(1)}(x) = \frac{e^2}{2}\int A_\nu(x')Tr\left\{\begin{array}{l}\gamma_\mu S^{ret}(x-x')\gamma_\nu S^1(x'-x) \\ +\gamma_\mu S^1(x-x')\gamma_\nu S^{adv}(x'-x)\end{array}\right\}d^4x' \tag{3.15}$$

where,

$$S_{\alpha\beta}^1(x-x') = -\langle 0|\left[\hat{\psi}_\alpha(x), \hat{\bar{\psi}}_\beta(x')\right]|0\rangle \tag{3.16}$$

This can be shown to be,

$$S_{\alpha\beta}^1(x-x') = \left(\gamma\frac{\partial}{\partial x} - m\right)\Delta^1(x-x'); \ \Delta^1(\vec{x},t) = \left(\frac{1}{2\pi}\right)^3 \int \exp[i\vec{k}\cdot\vec{x}]\frac{\cos\omega t}{\omega}d^3k \tag{3.17}$$

Pauli [11] shows that (3.15) can be written as,

$$j_{D\mu}^{(1)}(x) = e^2\int\left(K_{uv}(x-x') - \bar{K}_{uv}(x-x')\right)A_\nu(x')d^4x' \tag{3.18}$$

where,

$$K_{uv}(x) = 2\left\{\partial_\mu\Delta\partial_\nu\Delta^1 + \partial_\mu\Delta^1\partial_\nu\Delta - \delta_{uv}\left(\partial_\alpha\Delta\partial_\alpha\Delta^1 + m^2\Delta\Delta^1\right)\right\} \tag{3.19}$$

$$\bar{K}_{uv}(x-x') = -\varepsilon(t-t')K_{uv}(x-x'); \ \varepsilon(t) = \begin{cases}+1 & t>0 \\ -1 & t<0\end{cases} \tag{3.20}$$

$\bar{K}_{uv}(x-x')$ can also be expressed as,

$$\bar{K}_{\mu\nu}(x) = 4\left[\partial_\mu \bar{\Delta}(x)\partial_\nu \Delta^1(x) + \partial_\nu \bar{\Delta}(x)\partial_\mu \Delta^1(x) - \delta_{\mu\nu}\left(\partial_\rho \bar{\Delta}(x)\partial_\rho \Delta^1(x) + m^2 \bar{\Delta}(x)\Delta^1(x)\right)\right] \quad (3.21)$$

where,

$$\bar{\Delta}(x) = -\frac{1}{2}\varepsilon(t)\Delta(x) \quad (3.22)$$

The quantities $\Delta$, $\Delta^1$, and $\bar{\Delta}$ satisfy the following relationships,

$$\left(\Box - m^2\right)\Delta^1 = \left(\Box - m^2\right)\Delta = 0; \quad \left(\Box - m^2\right)\bar{\Delta} = -\delta^4(x); \quad \Box \equiv \partial_\rho \partial_\rho \quad (3.23)$$

Note that $\partial_\mu \bar{\Delta}(x-x')\partial_\nu \Delta^1(x-x') = \partial'_\mu \bar{\Delta}(x-x')\partial'_\nu \Delta^1(x-x')$ where $\partial'_\mu = \partial/\partial x'_\mu$.

Substitute $A_\nu(x) = \partial_\nu \chi(x)$ into (3.18) and integrate by parts to obtain,

$$j_{D\mu}^{(1)}(x) = -e^2 \int \chi(x')\partial'_\nu \left(K_{\mu\nu}(x-x') - \bar{K}_{\mu\nu}(x-x')\right)d^4x' \quad (3.24)$$

In order for the theory to be gauge invariant this equation must equal zero. For this to happen we must have $\partial_\nu \left(K_{\mu\nu}(x) - \bar{K}_{\mu\nu}(x)\right) = 0$. It can be shown that $\partial_\nu K_{\mu\nu}(x) = 0$. Therefore gauge invariance requires that $\partial_\nu \bar{K}_{\mu\nu}(x) = 0$. According to Pauli this is not the case. Pauli obtains,

$$\frac{\partial \bar{K}_{\mu\nu}}{\partial x_\nu} = -4\delta^4(x)\frac{\partial \Delta^1(x)}{\partial x_\mu} \quad (3.25)$$

Pauli states that this expression is not zero because of the highly divergent nature of $\Delta^1(x)$. Therefore the first order expansion of the vacuum current is not gauge invariant.

Notice that we have performed two different calculations of the vacuum current due to the presence of the potential $A_\nu(x) = \partial_\nu \chi(x)$. In the first calculation (Eqs. (3.4) through (3.6)) it was shown that the current operator is not affected by this potential which is consistent with a gauge invariant theory. In the second calculation (Eqs. (3.7) through (3.25)) it was shown that the vacuum current is changed by this potential (because $\partial_\nu \bar{K}_{\mu\nu}(x) \neq 0$) which is not consistent with a gauge invariant theory. So what went wrong? Why do two methods of calculation produce different results? I consider this an indication that the theory is either mathematically inconsistent in some way or an error was performed in Pauli's calculations leading up to the second result. The point of view will be explored further in the next section.

## 4. Expressing the vacuum current in manifestly gauge invariant form.

In order to better understand this problem I will attempt to write the expression for the vacuum current $j_{D\mu}^{(1)}(x)$ in a form that is manifestly gauge invariant. This will be the case if the electric potential terms $A_\nu$ always appear in the form $F_{\mu\nu} = \partial_\mu A_\nu - \partial_\nu A_\mu$. To proceed define the quantities,

$$I_\mu(x) = \int K_{\mu\nu}(x-x')A_\nu(x')d^4x'; \quad \bar{I}_\mu(x) = \int \bar{K}_{\mu\nu}(x-x')A_\nu(x')d^4x' \quad (4.1)$$

Therefore,

$$j_{D\mu}^{(1)}(x) = e^2\left(I_\mu(x) - \bar{I}_\mu(x)\right) \quad (4.2)$$

First examine $\bar{I}_\mu(x)$. Use (3.21) in the (4.1) to obtain,

$$\bar{I}_\mu(x) = 4\int \begin{bmatrix} \partial'_\mu \bar{\Delta}(x-x')\partial'_\nu \Delta^1(x-x') + \partial'_\nu \bar{\Delta}(x-x')\partial'_\mu \Delta^1(x-x') \\ -\delta_{\mu\nu}\left(\partial'_\rho \bar{\Delta}(x-x')\partial'_\rho \Delta^1(x-x') + m^2 \bar{\Delta}(x-x')\Delta^1(x-x')\right) \end{bmatrix} A_\nu(x')d^4x' \quad (4.3)$$

Through repeated use of integration by parts and Eq. (3.23) it is shown in Appendix 1 that this becomes,

$$\bar{I}_\mu(x) = 4\int \begin{Bmatrix} -2\left(\bar{\Delta}(x-x')\partial'_\nu \partial'_\mu \Delta^1(x-x')\right)\left(\frac{1}{\Box'}\partial'_\rho F_{\rho\nu}(x')\right) \\ +\left(\bar{\Delta}(x-x')\partial'_\nu \Delta^1(x-x')\right)F_{\nu\mu}(x') \\ -\left[(\Box'-m^2)\bar{\Delta}(x-x')\right]\left(\partial'_\mu \Delta^1(x-x')\right)\frac{\partial'_\nu}{\Box'}A_\nu(x') \end{Bmatrix} d^4x' \quad (4.4)$$

Next consider $I_\mu(x)$. $I_\mu(x)$ can be derived from $\bar{I}_\mu(x)$, as given in Eq. (4.3), if $\bar{\Delta}(x-x')$ is replaced by $\Delta(x-x')/2$. Therefore $I_\mu(x)$ is equivalent to (4.4) with $\bar{\Delta}(x-x')$ replaced by $\Delta(x-x')$. Use this fact and $(\Box'-m^2)\Delta(x-x') = 0$ to obtain,

$$I_\mu(x) = 2\int \left\{-2\left(\Delta(x-x')\partial'_\nu \partial'_\mu \Delta^1(x-x')\right)\left(\frac{1}{\Box'}\partial'_\rho F_{\rho\nu}(x')\right) + \left(\Delta(x-x')\partial'_\nu \Delta^1(x-x')\right)F_{\nu\mu}(x')\right\} d^4x' \quad (4.5)$$

From the above we see that $I_\mu(x)$ is manifestly gauge invariant because the dependence on the electric potential is contained in terms of the form $F_{\nu\mu}$. However if we examine (4.4) it seen that this is not the case for $\bar{I}_\mu(x)$. It is evident that there is a term that appears to be non-gauge invariant. This term is,

$$\bar{I}_{\mu,NG}(x) \equiv 4\int \left[\left(-\left[(\Box'-m^2)\bar{\Delta}(x-x')\right]\left(\partial'_\mu \Delta^1(x-x')\right)\frac{\partial'_\nu}{\Box'}A_\nu(x')\right)\right]d^4x' \quad (4.6)$$

If we set $A_\nu(x') = \partial'_\nu \chi(x')$ in the above and use $\partial'_\nu \partial'_\nu = \Box'$ we obtain,

$$\bar{I}_{\mu,NG}(x) = 4\int \left[\left(-\left[(\Box'-m^2)\bar{\Delta}(x-x')\right]\left(\partial'_\mu \Delta^1(x-x')\right)\chi(x')\right)\right]d^4x' \quad (4.7)$$

This quantity must be zero if the theory is gauge invariant. Since $\chi(x')$ is an arbitrary function this means that the term $\left[(\Box'-m^2)\bar{\Delta}(x-x')\right]\left(\partial'_\mu \Delta^1(x-x')\right)$ must be zero. Use $\left[(\Box'-m^2)\bar{\Delta}(x-x')\right] = -\delta^4(x-x')$ to obtain,

$$\left[(\Box'-m^2)\bar{\Delta}(x-x')\right]\left(\partial'_\mu \Delta^1(x-x')\right) = -\delta^4(x-x')\left(\partial'_\mu \Delta^1(x-x')\right) \quad (4.8)$$

Comparing this expression to (3.25) we see have run into the same problem that Pauli encountered when trying to determine if that theory is gauge invariant.

At this point I will focus the discussion on whether or not (4.8) is zero and if it is not how to modify the theory so that it is. Too start this analysis I will do a detailed examination of,

$$B_\mu(x) \equiv \left[(\Box-m^2)\bar{\Delta}(x)\right]\left(\partial_\mu \Delta^1(x)\right) \quad (4.9)$$

which must be zero for the theory to be gauge invariant. Use $\partial \varepsilon(t)/\partial t = 2\delta(t)$ along with (3.22) and (3.23) to obtain,

$$\left(\Box - m^2\right)\bar{\Delta}(\vec{x},t) = \frac{\partial \delta(t)}{\partial t}\Delta(\vec{x},t) + 2\delta(t)\frac{\partial \Delta(\vec{x},t)}{\partial t} \tag{4.10}$$

To evaluate this refer to (3.14) and use,

$$\frac{\partial \delta(t)}{\partial t}\sin \omega t + 2\delta(t)\frac{\partial \sin \omega t}{\partial t} = \omega \delta(t)\cos \omega t = \omega \delta(t) \tag{4.11}$$

to obtain,

$$\left(\Box - m^2\right)\bar{\Delta}(\vec{x},t) = -\left(\frac{1}{2\pi}\right)^3 \delta(t) \int \exp\left[i\vec{k}\cdot\vec{x}\right]d^3k \tag{4.12}$$

Next evaluate $B_i(x)$ where $i = 1,2,3$. Use (3.17) along with $\delta(t)\cos \omega t = \delta(t)$ in (4.9).

$$B_i(x) = -\delta(t)\left(\frac{1}{2\pi}\right)^6 \int \exp\left[i\vec{k}\cdot\vec{x}\right]d^3k \left(\int \frac{ik_i}{\omega}\exp\left[i\vec{k}\cdot\vec{x}\right]d^3k\right) \tag{4.13}$$

So the question is – does this equal zero? It will simplify the discussion at this point to change from 3+1 dimensions to 1+1 dimensions where $z$ will be the space dimension. In 1+1 dimensions this becomes,

$$B_i(z,t;L,L') = -\delta(t)\left(\frac{1}{2\pi}\right)^2 \left(\int_{-L'}^{+L'} e^{ik'z}dk'\right)\left(\int_{-L}^{+L} \frac{ik}{\omega}e^{ikz}dk\right) \tag{4.14}$$

Note that I have put in limits of integration, $\pm L$ and $\pm L'$, where $L \to \infty$ and $L' \to \infty$. This is necessary in order to evaluate the integrals. Take the Fourier transform of the above to obtain,

$$\tilde{B}_i(p,t;L,L') = \int B_i(z,t;L,L')e^{-ipz}dz = -\delta(t)\left(\frac{1}{(2\pi)}\right)^2 \left(\int_{-L}^{+L} \frac{ik}{\omega}dk \int_{-L'}^{+L'} \delta(k'+k-p)dk'\right) \tag{4.15}$$

Note that,

$$\int_{-L'}^{+L'} \delta(k'+k-p)dk' = \theta(L' - |k-p|) \tag{4.16}$$

Use this in (4.15) to obtain,

$$\tilde{B}_i(p,t;L,L') = -\delta(t)\left(\frac{1}{(2\pi)}\right)^2 \left(\int_{-L}^{+L} \frac{ik}{\omega}\theta(L' - |k-p|)dk\right) \tag{4.17}$$

Gauge invariance requires that $B_i(z,t;L,L')$ equals zero which means that $\tilde{B}_i(p,t;L,L') = 0$. Let's see if this is true. Evaluate $\tilde{B}_i(p,t;L,L')$ for the case $L = L'$ and assume $p > 0$ to obtain,

$$\tilde{B}_i(p,t;L,L) = -\delta(t)\left(\frac{1}{(2\pi)}\right)^2 \left(\int_{-L}^{+L} \frac{ik}{\omega}dk\,\theta(L - |k-p|)\right) = -\delta(t)\left(\frac{1}{(2\pi)}\right)^2 \left(\int_{-L+p}^{+L} \frac{ik}{\omega}dk\right) \tag{4.18}$$

To evaluate this use,

$$\left( \int_{-L+p}^{+L} \frac{ik}{\omega} dk \right) = \frac{i}{2} \Big|_{-L+p}^{+L} \sqrt{k^2 + m^2} = \frac{i}{2} \left( \sqrt{L^2 + m^2} - \sqrt{(-L+p)^2 + m^2} \right) \xrightarrow{L \to \infty} \frac{i}{2} p \qquad (4.19)$$

Therefore,

$$\tilde{B}_i(p, t; L, L) = -\delta(t) \left( \frac{1}{(2\pi)} \right)^2 \frac{i}{2} p \qquad (4.20)$$

In order for the theory to be gauge invariant we must have $\tilde{B}_i(p, t; L, L') = 0$, therefore this result seems to show that the theory is not gauge invariant. However there is a way to correct this problem. Go back to (4.17) and let $L' \gg |L \pm p|$. In this case $\theta(L' - |k - p|) = 1$ for all $k$ within the limits of integration $-L$ and $+L$. In this case (4.17) becomes,

$$\tilde{B}_i(p, t; L, L') \Big|_{L' \gg L} = -\delta(t) \left( \frac{1}{(2\pi)} \right)^2 \left( \int_{-L}^{+L} \frac{ik}{\omega} dk \right) = 0 \qquad (4.21)$$

Therefore if the limits of integration are correctly specified we can obtain a gauge invariant result. The next step is to figure out how to put these correct limits of integration into the beginning of the calculation so that they will automatically appear at the end. First, since $L' \gg L$ we can replace $L'$ with $\infty$ in (4.14). Next we can also replace limit of integration $\pm L$ with $\pm \infty$ provided we add $\theta(L - |k|)$ to the integrand. The result is,

$$B_i(z, t; L) = -\delta(t) \left( \frac{1}{(2\pi)} \right)^2 \left( \int_{-\infty}^{+\infty} e^{ik'z} dk' \right) \left( \int_{-\infty}^{+\infty} \frac{ik}{\omega} e^{ikz} \theta(L - |k|) dk \right) = 0 \qquad (4.22)$$

where the dependence $B_i$ on $L$ is explicitly shown. $L \to \infty$ but is a retained in the calculation as a finite, but arbitrary large positive number. Recall that this result applies to the 1+1 dimensional space. In 3+1 dimensions we have,

$$B_i(x; L) = -\delta(t) \left( \frac{1}{(2\pi)^3} \right)^2 \left( \int e^{i\vec{k}' \cdot \vec{x}} d^3k' \right) \left( \int \frac{ik_i}{\omega} e^{i\vec{k} \cdot \vec{x}} \theta(L - |\vec{k}|) d^3k \right) = 0 \qquad (4.23)$$

At this point it and been shown that we achieve a gauge invariant result if we can add the cutoff factor $\theta(L - |\vec{k}|)$ into the expression for $\Delta^1(x)$. We define $\Delta_L^1(x)$ by,

$$\Delta_L^1(\vec{x}, t) = \left( \frac{1}{2\pi} \right)^3 \int \exp[i\vec{k} \cdot \vec{x}] \frac{\cos \omega t}{\omega} \theta(L - |\vec{k}|) d^3k \qquad (4.24)$$

If we replace $\Delta^1(x)$ by $\Delta_L^1(x)$ in Eqs. (3.19) and (3.21) then the result will be gauge invariant. Now where does $\Delta^1$ come from? It comes from Eq. (3.17) which is a mathematical evaluation of $S^1(x)$ which in turn is defined in Eq. (3.16). We can, then, achieve a gauge invariant theory by modifying the definition of the field operator. Define the modified field operator by,

$$\hat{\psi}_\alpha^{(0)}(x;L) = \frac{1}{\sqrt{V}} \sum_{\vec{q}} \theta(L-|\vec{q}|) \left[ e^{i(\vec{q}\cdot\vec{x}-Et)} \sum_{r=1}^{2} \hat{a}_r(\vec{q}) u_\alpha^{(r)}(\vec{q}) + e^{i(\vec{q}\cdot\vec{x}+Et)} \sum_{r=3}^{4} \hat{a}_r(\vec{q}) u_\alpha^{(r)}(\vec{q}) \right] \quad (4.25)$$

where $L \to \infty$. When this is done then $S^1(x)$ becomes,

$$S_{\alpha\beta}^1(x-x';L) = \left( \gamma \frac{\partial}{\partial x} - m \right) \Delta_L^1(x-x') \quad (4.26)$$

The result of these changes is that $\Delta_L^1(x)$ instead of $\Delta^1(x)$ will appear in Eq. (4.4). The result of this is that non-gauge invariant term, $\bar{I}_{\mu,NG}(x)$, given by (4.6), will be zero.

In this case (4.4) becomes,

$$\bar{I}_{\mu,vac}(x;L) = 4\int \left[ \begin{array}{l} -2(\bar{\Delta}(x-x')\partial_\nu'\partial_\mu'\Delta_L^1(x-x'))\left(\frac{1}{\Box'}\partial_\rho' F_{\rho\nu}(x')\right) \\ +(\bar{\Delta}(x-x')\partial_\nu'\Delta_L^1(x-x')) F_{\nu\mu}(x') \end{array} \right] d^4x' \quad (4.27)$$

## 5. Discussion and Further analysis.

It was shown in the last section there is an exact solution to the Dirac equation where $A_\nu(x) = \partial_\nu \chi(x)$ and that the current operator is invariant with respect to this solution. This is consistent with a gauge invariant theory. However when perturbation theory is used it is found that if $A_\nu(x) = \partial_\nu \chi(x)$ there will be a non-gauge invariant term. This term can be eliminated if the quantity $\Delta^1(x)$ is redefined as the quantity $\Delta_L^1(x)$. This, in turn led to a redefining of the field operator from $\hat{\psi}_\alpha^{(0)}(x)$ to $\hat{\psi}_\alpha^{(0)}(x;L)$. The problem with this analysis, in my opinion, is that we had to get to the end of a long detailed calculation to realize that we had made a "mistake" that required correction. I will argue that is should have been possible to have detected this problem earlier on in the calculation. I will show that the mistake occurs in the way the first order term $e\hat{\psi}^{(1)}(x)$ is defined in Eq. (3.11) which I reproduce below for convenience,

$$e\hat{\psi}^{(1)}(x) = -ie\int S^{ret}(x-x')\gamma_\nu A_\nu(x')\hat{\psi}^{(0)}(x') d^4x' \quad (3.11)$$

We have already shown that if $A_\nu(x) = \partial_\nu \chi(x)$ then $\hat{\psi}(x) = e^{ie\chi(x)}\hat{\psi}^{(0)}(x)$. Expand the exponential to obtain $\hat{\psi}(x) = (1+ie\chi(x)+\ldots)\hat{\psi}^{(0)}(x)$. Therefore if $A_\nu(x) = \partial_\nu \chi(x)$ is substituted into (3.11) we should obtain $e\hat{\psi}^{(1)}(x) = ie\chi(x)\hat{\psi}^{(0)}(x)$. Let's see if this is true. Substitute $A_\nu(x) = \partial_\nu \chi(x)$ into (3.11) and integrate by parts to obtain,

$$e\hat{\psi}^{(1)}(x) = ie\int \chi(x')\partial_\nu'\left[ S^{ret}(x-x')\gamma_\nu \hat{\psi}^{(0)}(x') \right] d^4x' \quad (5.1)$$

Use $S^{ret}(x-x') = -\theta(t-t')S(x-x')$ and $\partial\theta(t-t')/\partial t' = -\delta(t-t')$ to obtain,

$$e\hat{\psi}^{(1)}(x) = -ie\int \chi(x') \left[ \begin{array}{l} \theta(t-t')\partial_\nu'\left( S(x-x')\gamma_\nu \hat{\psi}^{(0)}(x') \right) \\ -\delta(t-t')S(x-x')\gamma_0 \hat{\psi}^{(0)}(x') \end{array} \right] d^4x' \quad (5.2)$$

It can be shown that $\partial'_\nu (S(x-x')\gamma_\nu \hat{\psi}^{(0)}(x')) = 0$. Also, for convenience, set $t = 0$ to obtain,

$$e\hat{\psi}^{(1)}(\vec{x},0) = ie\int \chi(\vec{x}',0)\left[S(\vec{x}-\vec{x}',0)\gamma_0 \hat{\psi}^{(0)}(\vec{x}',0)\right]d^3x' \tag{5.3}$$

Using (3.14) it can be shown that,

$$S(\vec{x},0) = \gamma_0 \int e^{i\vec{k}\cdot\vec{x}} \frac{d^3k}{(2\pi)^3} \tag{5.4}$$

The field operator $\hat{\psi}^{(0)}(x)$ was given in Eq. (3.2) as an infinite sum. Here we shall rewrite it as an integral to obtain,

$$\hat{\psi}_\alpha^{(0)}(\vec{x},0) = \left(\frac{1}{2\pi}\right)^3 \sum_{r=1}^{4}\int d^3q\, e^{i\vec{q}\cdot\vec{x}} \hat{a}_r(\vec{q})u_\alpha^{(r)}(\vec{q}) \tag{5.5}$$

Use all this in (5.3) to obtain to obtain,

$$e\hat{\psi}^{(1)}(\vec{x},0) = ie\sum_{r=1}^{4}\int d^3x' \left[\chi(\vec{x}',0)\left(\int e^{i\vec{k}\cdot(\vec{x}-\vec{x}')}\frac{d^3k}{(2\pi)^3}\right)\left(\int \frac{d^3q}{(2\pi)^3}e^{i\vec{q}\cdot\vec{x}}\hat{a}_r(\vec{q})u_\alpha^{(r)}(\vec{q})\right)\right] \tag{5.6}$$

For the moment drop the requirement that $\chi(\vec{x},0)$ is a real valued function and set $\chi(\vec{x},0) = e^{-i\vec{p}\cdot\vec{x}}$. Next perform the spatial integration to obtain,

$$e\hat{\psi}^{(1)}(\vec{x},0) = ie\sum_{r=1}^{4}\left\{\int \frac{d^3q}{(2\pi)^3}\hat{a}_{\vec{q},r}u(\vec{q},r)\left(\int \frac{d^3k}{(2\pi)^3}e^{i\vec{k}\cdot\vec{x}}\delta^3(\vec{q}-\vec{k}-\vec{p})\right)\right\} \tag{5.7}$$

As before we will simply this expression by converting from three to one space dimension and add limits of integration. For this problem we obtain,

$$e\hat{\psi}^{(1)}(z,0;L,L') = ie\sum_{r=1}^{4}\left\{\int_{-L}^{+L} \frac{dq}{(2\pi)}\hat{a}_{q,r}u(q,r)\left(\int_{-L'}^{+L'} \frac{dk}{(2\pi)}e^{ik\cdot z}\delta^3(q-k-p)\right)\right\} \tag{5.8}$$

where the limits $L, L' \to \infty$. Use the following relationship to obtain,

$$\int_{-L'}^{+L'} \frac{dk}{(2\pi)}e^{ik\cdot z}\delta^3(q-k-p) = \theta(L'-|q-p|)e^{i(q-p)z} \tag{5.9}$$

Use this in (5.8) to obtain,

$$e\hat{\psi}^{(1)}(z,0;L,L') = ie\sum_{r=1}^{4}\left\{\int_{-L}^{+L} \frac{dq}{(2\pi)}\hat{a}_{q,r}u(q,r)\theta(L'-|q-p|)e^{i(q-p)z}\right\} \tag{5.10}$$

Evaluate this for $L = L'$ and assume $p > 0$ to obtain,

$$e\hat{\psi}^{(1)}(z,0;L,L) = \sum_{r=1}^{4}\left\{\int_{-L+p}^{+L} \frac{dq}{(2\pi)}\hat{a}_{q,r}u(q,r)e^{i(q-p)z}\right\} = (ie)e^{-ipz}\sum_{r=1}^{4}\left\{\int_{-L+p}^{+L} \frac{dq}{(2\pi)}\hat{a}_{q,r}u(q,r)e^{iqz}\right\} \tag{5.11}$$

For this one space dimensional example,

$$\chi(z,0) = e^{-ipz} \text{ and } e\hat{\psi}^{(0)}(z,0;L) = (ie)\sum_{r=1}^{4}\left\{\int_{-L}^{+L} \frac{dq}{(2\pi)}\hat{a}_{q,r}u(q,r)e^{iqz}\right\} \tag{5.12}$$

Recall that for a gauge invariant theory we want $e\hat{\psi}^{(1)}(z,0;L,L')$ to equal $ie\chi(z,0)\hat{\psi}^{(0)}(z,0;L)$. If we examine (5.11) and (5.12) we see that this is not the case. The problem is due to the lower limit of integration in (5.11) which is $(-L+p)$ instead of $-L$.

The solution to this problem is to specify that $L' \gg L$. When this is done then $\theta(L'-|q-p|)$ in the integrand of (5.11) can be replaced by 1. Therefore (5.10) becomes,

$$e\hat{\psi}^{(1)}(z,0;L,L') \underset{L' \gg L}{=} e^{-ipz}(ie)\sum_{r=1}^{4}\left\{\int_{-L}^{+L}\frac{dq}{(2\pi)}\hat{a}_{q,r}u(q,r)e^{iqz}\right\} \quad (5.13)$$

When this result is compared with (5.12) we obtain,

$$e\hat{\psi}^{(1)}(z,0;L,L') \underset{L' \gg L}{=} ie\chi(z,0)\hat{\psi}^{(0)}(z,0;L) \quad (5.14)$$

Therefore we see in order to correctly do the perturbative expansion the limits of integration must be correctly chosen. In our one dimensional example this is achieved by making sure that $L' \gg L$. In 3+1 dimensions this is ensured by modifying the field operator by adding the factor $\theta(L-|\vec{q}|)$ into the definition of the state vector so that (5.5) becomes,

$$\hat{\psi}^{(0)}(\vec{x},0) = \sum_{r=1}^{4}\int \hat{a}_{\vec{q},r}u(\vec{q},r)\theta(L-|\vec{q}|)e^{+i\vec{q}\cdot\vec{x}}\frac{d^3q}{(2\pi)^3} \quad (5.15)$$

where $L \to \infty$. In other words for the expansion in perturbation theory to give the correct result the field operator must be defined using a cutoff function.

## 6. Polarization tensor in momentum space.

In this section we will examine the vacuum polarization tensor in momentum space. We could use the results obtained above however it will be more informative to use the expression for the vacuum polarization tensor given by G. Kallen [12]. According to Kallen the vacuum polarization tensor is,

$$\Pi_{\mu\nu}(p) = \frac{e^2}{16\pi^2}\iint d^4k'd^4k''\left\{\begin{array}{l}\delta(p-k'+k'')Tr[\gamma_\mu(i\gamma k'-m)\gamma_\nu(i\gamma k''-m)]\\ \times[\delta(k'^2+m^2)f(k'') + \delta(k''^2+m^2)f^*(k')]\end{array}\right\} \quad (6.1)$$

where,

$$f(k) = P\frac{1}{k^2+m^2} - i\pi\varepsilon(k_0)\delta(k^2+m^2) \quad (6.2)$$

Kallen shows that this expression is not gauge invariant. However based on the previous discussion we know that in order to achieve a gauge invariant theory we have to modify the field operator by including a cutoff per Eq. (4.25). When this is done (6.1) becomes,

$$\Pi_{\mu\nu}(p;L) = \frac{e^2}{16\pi^3}\iint d^4k'd^4k''\left\{\begin{array}{l}\delta(p-k'+k'')Tr[\gamma_\mu(i\gamma k'-m)\gamma_\nu(i\gamma k''-m)]\\ \times[\delta(k'^2+m^2)f(k'')\theta(L-|\vec{k}'|) + \delta(k''^2+m^2)f^*(k')\theta(L-|\vec{k}''|)]\end{array}\right\}$$

(6.3)

As will be shown this will result in an expression that is manifestly gauge invariant. This, of course, differs from Kallen who showed that $\Pi_{\mu\nu}(p)$ was not gauge invariant (See discussion in section 30 of [12]).

Evaluate the trace as follows,

$$Tr\left[\gamma_\mu(i\gamma k'-m)\gamma_\nu(i\gamma k''-m)\right] = -4\left[k'_\mu k''_\nu + k'_\nu k''_\mu - \delta_{\mu\nu}(k'k''+m^2)\right] \quad (6.4)$$

Use this in (6.3) and relabel dummy indices to obtain,

$$\Pi_{\mu\nu}(p,L) = \frac{-4e^2}{16\pi^3}\iint d^4k'd^4k'' \left\{\begin{bmatrix}\left[k'_\mu k''_\nu + k'_\nu k''_\mu - \delta_{\mu\nu}(k'k''+m^2)\right] \\ \times \delta(k'^2+m^2)\theta(L-|\vec{k}'|)\end{bmatrix}\begin{Bmatrix}\delta(p-k'+k'')f(k'') \\ +\left[\delta(p-k''+k')f^*(k'')\right]\end{Bmatrix}\right\} \quad (6.5)$$

Next use $f^*(-k'') = f(k'')$ to obtain

$$\Pi_{\mu\nu}(p,L) = \frac{-e^2}{2\pi^3}\iint d^4k'd^4k''\left\{\delta(p-k'+k'')\begin{bmatrix}k'_\mu k''_\nu + k'_\nu k''_\mu \\ -\delta_{\mu\nu}(k'k''+m^2)\end{bmatrix}\delta(k'^2+m^2)f(k'')\theta(L-|\vec{k}'|)\right\} \quad (6.6)$$

Integrate with respect to $k''$ to obtain,

$$\Pi_{\mu\nu}(p,L) = \frac{-e^2}{2\pi^3}\int d^4k'\left\{\begin{bmatrix}k'_\mu(k'_\nu-p_\nu)+k'_\nu(k'_\mu-p_\mu) \\ -\delta_{\mu\nu}(k'(k'-p)+m^2)\end{bmatrix}\delta(k'^2+m^2)f(k'-p)\theta(L-|\vec{k}'|)\right\} \quad (6.7)$$

Rearrange terms and relabel the dummy indices to obtain,

$$\Pi_{\mu\nu}(p,L) = \frac{-e^2}{2\pi^3}\int d^4k\left\{\begin{bmatrix}2k_\mu k_\nu - k_\mu p_\nu - k_\nu p_\mu \\ -\delta_{\mu\nu}(k^2-kp+m^2)\end{bmatrix}\delta(k^2+m^2)f(k-p)\theta(L-|\vec{k}|)\right\} \quad (6.8)$$

The current in momentum space is given by,

$$j^{(1)}_{D\mu}(p;L) = \Pi_{\mu\nu}(p,L)A_\nu(p) \quad (6.9)$$

To show that this is gauge invariant we will put $\Pi_{\mu\nu}(p,L)A_\nu(p)$ in a form that is manifestly gauge invariant. To do this I want $A_\nu(p)$ to always appear in the form $\tilde{F}_{\mu\nu}(p) = i(p_\mu A_\nu - p_\nu A_\mu)$. It is shown in Appendix 2 that,

$$j^{(1)}_{D\mu}(p;L) = \frac{-e^2}{2\pi^3}(-i\tilde{F}_{\nu\rho})\int d^4k\left\{\left[\frac{2k_\mu k_\rho p_\nu}{p^2}+k_\nu\delta_{\rho\mu}\right]\delta(k^2+m^2)f(k-p)\theta(L-|\vec{k}|)\right\} \quad (6.10)$$

which is manifestly gauge invariant. Use (6.2) in the above to obtain,

$$j^{(1)}_{D\mu}(p;L) = j_{DAu}(p;L) + j_{DBu}(p;L) \quad (6.11)$$

$$j_{DAu}(p;L) = \frac{-e^2}{2\pi^3}(-i\tilde{F}_{\nu\rho})\int d^4k\left\{\left[\frac{2k_\mu k_\rho p_\nu}{p^2}+k_\nu\delta_{\rho\mu}\right]\delta(k^2+m^2)P\frac{1}{(k-p)^2+m^2}\theta(L-|\vec{k}|)\right\} \quad (6.12)$$

$$j_{DBu}(p;L) = \frac{-e^2}{2\pi^3}(-i\tilde{F}_{\nu\rho})\int d^4k\left\{\left[\frac{2k_\mu k_\rho p_\nu}{p^2}+k_\nu\delta_{\rho\mu}\right]\delta(k^2+m^2)\begin{pmatrix}-i\pi\varepsilon(k_0-p_0)\\ \times\delta((k-p)^2+m^2)\end{pmatrix}\theta(L-|\vec{k}|)\right\} \quad (6.13)$$

## 7. Evaluation of the vacuum current.

If $|p_0| > |\vec{p}|$ we can always do a Lorentz boost to a reference frame where $\vec{p} = 0$. In this section we will evaluate the vacuum current for this case. Therefore the electric potential has no spatial dependence and is only dependent on time. We will evaluate (6.10) assume the following: $A_0 = 0$, $\vec{A}(\vec{x},t) = (A_1(t),0,0)$, $\vec{p} = 0$, and $2m > |p_0| > 0$. This last assumption will mean the denominator in the integrand will never be zero so that we can drop the $P$ (indicating "principle part"). With this assumptions the only non-zero terms for $\tilde{F}_{vp}(p_4)$ is $\tilde{F}_{41}(p_4) = ip_4 A_1(p_4)$ and $\tilde{F}_{14}(p_4) = -ip_4 A_1(p_4)$ where $p_4 = ip_0$. For this problem it is shown in Appendix 3 that,

$$j_{DA1}(p_0;L) = \frac{e^2 A_1(p_0)}{6\pi^2} \int_{2m}^{\infty} dz \frac{(z^2 + 2m^2)\sqrt{z^2 - 4m^2}}{(z^2 - p_0^2)} \theta(L - |\vec{k}|) \tag{7.1}$$

where $|\vec{k}| = \sqrt{z^2/4 - m^2}$. Next use the following in the above,

$$\frac{1}{(z^2 - p_0^2)} = \left[\frac{1}{(z^2 - p_0^2)} - \frac{1}{z^2}\right] + \frac{1}{z^2} = \frac{p_0^2}{z^2(z^2 - p_0^2)} + \frac{1}{z^2} \tag{7.2}$$

to obtain,

$$j_{DA1}(p_0;L) = j_{DAA1}(p_0;L) + j_{DAB1}(p_0;L) \tag{7.3}$$

where,

$$j_{DAA1}(p_0;L) = \frac{e^2 p_0^2 A_1(p_0)}{6\pi^2} \int_{2m}^{\infty} dz \left(\frac{\sqrt{z^2 - 4m^2}(z^2 + 2m^2)}{z^2(z^2 - p_0^2)}\right) \theta(L - |\vec{k}|) \tag{7.4}$$

$$j_{DAB1}(p_0;L) = \frac{e^2 A_1(p_0)}{6\pi^2} \int_{2m}^{\infty} dz \frac{(z^2 + 2m^2)\sqrt{z^2 - 4m^2}}{z^2} \theta(L - |\vec{k}|) \tag{7.5}$$

In terms of the vacuum polarization tensor we use $j_{DAA1}(p_0;L) = \Pi_{DAA11}(p_0;L) A_1(p_0)$ and $j_{DAB1}(p_0;L) = \Pi_{DAB11}(p_0;L) A_1(p_0)$ so that the total polarization tensor for this example is $\Pi_{11}(p_0;L) = \Pi_{DAA11}(p_0;L) + \Pi_{DAB11}(p_0;L)$ where,

$$\Pi_{DAA11}(p_0;L) = \frac{e^2 p_0^2}{6\pi^2} \int_{2m}^{\infty} dz \left(\frac{\sqrt{z^2 - 4m^2}(z^2 + 2m^2)}{z^2(z^2 - p_0^2)}\right) \theta(L - |\vec{k}|) \tag{7.6}$$

and,

$$\Pi_{DAB11}(p_0;L) = \frac{e^2}{6\pi^2} \int_{2m}^{\infty} dz \frac{(z^2 + 2m^2)\sqrt{z^2 - 4m^2}}{z^2} \theta(L - |\vec{k}|) \tag{7.7}$$

Compare this with Heitler's results which are given in Eqs. (2.7) and (2.8). It can be readily shown that if we drop the term $\theta(L - |\vec{k}|)$ from the integrand that $\Pi_{DAA11}(p_0;L) = \pi_{11}^G(p_0)$ where $\pi_{11}^G(p_0)$ is obtained by setting $\mu = \nu = 1$ in (2.7) along with $\vec{p} = 0$. Now compare $\Pi_{DAB11}(p_0;L)$ to $\pi_{11}^{NG}(p_0)$. If we ignore the term $\theta(L-|\vec{k}|)$ the quantity in the integrands is the same. The only difference in the non-gauge invariant factor $\delta_{\mu\nu}(1-\delta_{\mu 4})$ that appeared in the term for $\pi_{\mu\nu}^{NG}(p_0)$. This term is what made Heitler's result non-gauge invariant. However in the approach taken here we have a fully gauge invariant theory. The result is that this non-gauge invariant term is now fully gauge invariant with $\delta_{\mu\nu}(1-\delta_{\mu 4})$ by 1. Thus the theory is mathematically consistent. However from the point of view of physics this creates a problem because the term $\Pi_{DAB11}(p_0;L)$ is highly divergent.

Next use (7.2) in (7.4) to obtain,

$$j_{DAA1}(p_0;L) = j_{DAAA1}(p_0;L) + j_{DAAB1}(p_0;L) \tag{7.8}$$

where,

$$j_{DAAA1}(p_0;L) = \frac{e^2 p_0^4 A_1(p_0)}{6\pi^2} \int_{2m}^{\infty} dz \left( \frac{(z^2+2m^2)\sqrt{z^2-4m^2}}{z^4(z^2-p_0^2)} \right) \tag{7.9}$$

$$j_{DAAB1}(p_0;L) = \frac{e^2 p_0^2 A_1(p_0)}{6\pi^2} \int_{2m}^{\infty} dz \left( \frac{(z^2+2m^2)\sqrt{z^2-4m^2}}{z^4} \right) \theta(L-|\vec{k}|) \tag{7.10}$$

Note that the cuttof term $\theta(L-|\vec{k}|)$ was dropped from Eq. (7.9). This is because the integrand converges fast enough that $\theta(L-|\vec{k}|)$ has no effect as $L \to \infty$. It is readily apparent that $j_{DAAB1}(p_0;L)$ is logarithmically divergent. However it can be readily shown that this is just a charge renormalization term because, from Maxwell's equations, $p_0^2 A_1(p_0)$ is proportional to the applied current. Finally $j_{DAAA1}(p_0;L)$ is finite and is the standard result for the vacuum current. Also it is shown Appendix 3 that,

$$j_{DB1}(p_0;L) = \frac{-i\pi e^2 |p_0| \tilde{F}_{41}}{12\pi^2} \sqrt{1-\frac{4m^2}{p_0^2}} \left( 1 + \frac{2m^2}{p_0^2} \right) \theta(|p_0|-2m) \tag{7.11}$$

However this is zero for the range $2m > |p_0| > 0$ so we will not discuss it further.

So at this point the total vacuum current for this example in the range $2m > |p_0| > 0$ is given by,

$$j_{D1}^{(1)}(p_0;L) = \left( j_{DAAA1}(p_0;L) + j_{DAAB1}(p_0;L) \right) + j_{DAB1}(p_0;L) \tag{7.12}$$

where $j_{DAAA1}(p_0;L)$ is the finite standard result, $j_{DAAB1}(p_0;L)$ is the logarithmically divergent charge renormalization term, and $j_{DAB1}(p_0;L)$ is a highly divergent term which must be eliminated. It will be shown that this term will be cancelled out if there are two scalar fields with the same mass as the Dirac field.

## 8. Polarization tensor for the scalar field.

In this section I will examine gauge invariance in scalar field theory. The vacuum polarization current for a charged scalar field is given in Section 9 of [11] as,

$$j^{(1)}_{S\mu}(x) = e^2 \int \left( L_{\mu\nu}(x-x') - \bar{L}_{\mu\nu}(x-x') \right) A_\nu(x') d^4x' \tag{8.1}$$

where,

$$\bar{L}_{\mu\nu}(x) = -\begin{bmatrix} \partial_\mu \bar{\Delta}(x) \partial_\nu \Delta^1_L(x) + \partial_\nu \bar{\Delta}(x) \partial_\mu \Delta^1_L(x) - \bar{\Delta}(x) \partial_\mu \partial_\nu \Delta^1_L(x) \\ -\Delta^1_L(x) \partial_\mu \partial_\nu \bar{\Delta}(x) - \delta_{\mu\nu} \delta^4(x) \Delta^1_L(x) \end{bmatrix} \tag{8.2}$$

and,

$$L_{\mu\nu}(x) = -\frac{1}{2} \left[ \partial_\mu \Delta(x) \partial_\nu \Delta^1_L(x) + \partial_\nu \Delta(x) \partial_\mu \Delta^1_L(x) - \Delta(x) \partial_\mu \partial_\nu \Delta^1_L(x) - \Delta^1_L(x) \partial_\mu \partial_\nu \Delta(x) \right] \tag{8.3}$$

Note the in the above expressions I have replaced $\Delta^1$ with $\Delta^1_L(x)$. Using integration by parts it is shown in Appendix 4 that scalar vacuum current can be expressed as,

$$j^{(1)}_{S\mu}(x) = e^2 \int \left\{ -\frac{1}{2} \begin{pmatrix} K_{\mu\nu}(x-x') \\ -\bar{K}_{\mu\nu}(x-x') \end{pmatrix} A_\nu(x') + \begin{pmatrix} \frac{1}{2}\Delta(x-x') \\ -\bar{\Delta}(x-x') \end{pmatrix} \Delta^1_L(x-x') \partial_\nu F_{\mu\nu}(x') \right\} d^4x' \tag{8.4}$$

Next, refer to (3.18) to obtain,

$$j^{(1)}_{S\mu}(x) = -\frac{1}{2} j^{(1)}_{D\mu}(x) + e^2 \int \left( \left( \frac{1}{2}\Delta(x-x') - \bar{\Delta}(x-x') \right) \Delta^1_L(x-x') \partial_\nu F_{\mu\nu}(x') \right) d^4x' \tag{8.5}$$

From Maxwell's equations $\partial_\nu F_{\mu\nu}(x) = j_{A\mu}(x)$ where $j_{A\mu}(x)$ is the applied current. Also define,

$$S_{\mu 1}(x;L) = \int \frac{1}{2} \Delta(x-x') \Delta^1_L(x-x') j_{A\mu}(x') d^4x' \tag{8.6}$$

and,

$$S_{\mu 2}(x;L) = \int \bar{\Delta}(x-x') \Delta^1_L(x-x') j_{A\mu}(x') d^4x' \tag{8.7}$$

Use all this in (8.5) to obtain,

$$j^{(1)}_{S\mu}(x) = -\frac{1}{2} j^{(1)}_{D\mu}(x) + e^2 \left( S_{\mu 1}(x;L) - S_{\mu 2}(x;L) \right) \tag{8.8}$$

In momentum space this becomes,

$$j^{(1)}_{S\mu}(p) = -\frac{1}{2} j^{(1)}_{D\mu}(p) + e^2 \left( S_{\mu 1}(p;L) - S_{\mu 2}(p;L) \right) \tag{8.9}$$

This will be evaluated for the case $\vec{p} = 0$, $2m > |p_0| > 0$. From Appendix 5,

$$S_{\mu 1}(p_0;L) = f_1(p_0) j_{A\mu}(p_0); \quad S_{\mu 2}(p_0;L) = \left( f_{2a}(p_0) + f_{2b}(p_0) \right) j_{A\mu}(p_0) \tag{8.10}$$

where,

$$f_1(p_0) = \frac{-ip_0}{16\pi}\sqrt{1 - \frac{4m^2}{p_0^2}}\theta(|p_0| - 2m) \tag{8.11}$$

$$f_{2a}(p) = -\frac{p_0^2}{8\pi^2}\int_{2m}^{\infty} dz \left[\frac{\sqrt{z^2 - 4m^2}}{z^2(z^2 - p_0^2)}\right]; \quad f_{2b}(p) = -\frac{1}{8\pi^2}\int_{2m}^{\infty} dz \left[\frac{\sqrt{z^2 - 4m^2}}{z^2}\right]\theta(L - |\vec{k}|) \tag{8.12}$$

Write,

$$j_{S\mu}^{(1)}(p_0) = j_{Sa\mu}(p_0) + j_{Sb\mu}(p_0) + j_{Sc\mu}(p_0) - j_{D\mu}^{(1)}(p_0)/2 \tag{8.13}$$

where,

$$j_{Sa\mu}(p_0) = -e^2 f_{2a}(p_0) j_{A\mu}(p_0); \quad j_{Sb\mu}(p_0) = -e^2 f_{2b}(p_0) j_{A\mu}(p_0); \quad j_{Sc\mu}(p_0) = e^2 f_1(p_0) j_{A\mu}(p_0) \tag{8.14}$$

For the range $2m > |p_0| > 0$ $j_{Sc\mu}(p_0)$ will be zero, $j_{Sb\mu}(p_0)$ is logarithmically divergent but is proportional to the applied current so is simply a charge renormalization term, and $j_{Sa\mu}(p_0)$ is finite. The divergent terms are all included in the $j_{D\mu}^{(1)}(p_0)/2$ term. Therefore as in the case of the Dirac field we have highly divergent terms which must be removed if the theory is to be consistent with the real world. To do this we can invoke a Supersymmetric solution. We assume that there are two charged scalar fields for every Dirac field and both fields have the same mass. In this case the total vacuum current is,

$$j_{T\mu}^{(1)}(p_0) = j_{D\mu}^{(1)}(p_0) + 2j_{S\mu}^{(1)}(p_0) = 2\left(j_{Sa\mu}(p_0) + j_{Sb\mu}(p_0) + j_{Sc\mu}(p_0)\right) \tag{8.15}$$

Therefore we are left with a logarithmically divergent charge renormalization terms and two finite terms, one of which is zero for the example $2m > |p_0| > 0$.

## 9. Summary and Conclusion.

In the beginning of this paper I noted that calculations of the vacuum current in quantum field theory produce non-gauge invariant results. In an attempt to understand why this problem exists I examined in detail a calculation of the vacuum current by Pauli. It was shown that this problem could be corrected by including a cutoff function in the definition of the field operator (see Eq. (4.25)). When this is done the vacuum current of the Dirac field will be gauge invariant. However there is a potential problem in that a divergent term which is nonphysical is included in the result. The divergent can be eliminated if the vacuum current due to a scalar field is included in the result. It was shown that if two scalar fields exist with the same mass as the Dirac field then this divergent term is eliminated.

## Appendix 1.

In the following we will rearrange terms in attempt to obtain an expression that is explicitly gauge invariant in order to derive Eq. (4.4). To simplify notation we will drop the explicit dependence on the coordinates and write $\bar{\Delta}$ and $\Delta^1$ for $\bar{\Delta}(x - x')$ and $\Delta^1(x - x')$, respectively. Also we will be repeatedly using integration by parts. In all cases the limits at infinity are assume to go to zero. In order to simplify notation instead of writing $\int u(\partial'_\nu w)d^4x' = -\int(\partial'_\nu u)w d^4x'$ I will write $u(\partial'_\nu w) \Rightarrow -(\partial'_\nu u)w$. First evaluate the following quantity.

$$\delta_{\mu\nu}\left(\partial'_\rho \bar{\Delta}\partial'_\rho \Delta^1 + m^2 \bar{\Delta}\Delta^1\right)A_\nu(x') = \left(\partial'_\rho \bar{\Delta}\partial'_\rho \Delta^1 + \bar{\Delta}\partial'_\rho \partial'_\rho \Delta^1\right)A_\mu(x') \tag{9.1}$$

where I have used (3.23). Integrate by parts to obtain,

$$\left(\partial'_\rho \bar{\Delta}\partial'_\rho \Delta^1 + \bar{\Delta}\partial'_\rho \partial'_\rho \Delta^1\right)A_\mu(x') \Rightarrow -\bar{\Delta}\partial'_\rho \Delta^1 \partial'_\rho A_\mu(x') \tag{9.2}$$

Next integrate by parts the term $\left(\partial'_\mu \bar{\Delta}\partial'_\nu \Delta^1\right)A_\nu(x')$ to obtain,

$$\left(\partial'_\mu \bar{\Delta}\partial'_\nu \Delta^1\right)A_\nu(x') \Rightarrow \left(-\bar{\Delta}\partial'_\mu \partial'_\nu \Delta^1\right)A_\nu(x') - \left(\bar{\Delta}\partial'_\nu \Delta^1\right)\partial'_\mu A_\nu(x') \tag{9.3}$$

Use this in (4.3) to obtain,

$$\bar{I}_\mu(x) = 4\int\left[\left(\partial'_\nu \bar{\Delta}\partial'_\mu \Delta^1 - \bar{\Delta}\partial'_\nu \partial'_\mu \Delta^1\right)A_\nu(x') + \bar{\Delta}\partial'_\nu \Delta^1\left(\partial'_\nu A_\mu(x') - \partial'_\mu A_\nu(x')\right)\right]d^4x'$$

Define,

$$F_{\nu\mu}(x') = \left(\partial'_\nu A_\mu(x') - \partial'_\mu A_\nu(x')\right); \quad S_{\mu\nu}(x-x') = \left(\partial'_\nu \bar{\Delta}\partial'_\mu \Delta^1 - \bar{\Delta}\partial'_\nu \partial'_\mu \Delta^1\right) \tag{9.4}$$

so that,

$$\bar{I}_\mu(x) = 4\int\left[S_{\mu\nu}A_\nu(x') + \left(\bar{\Delta}\partial'_\nu \Delta^1\right)F_{\nu\mu}(x')\right]d^4x' \tag{9.5}$$

Perform the following sequence of mathematic steps,

$$S_{\mu\nu}(x-x')A_\nu(x') = \left(\frac{1}{\Box}\right)\Box\left(S_{\mu\nu}(x-x')A_\nu(x')\right) = \left(\frac{1}{\Box}\right)\left(\Box' S_{\mu\nu}(x-x')A_\nu(x')\right) \tag{9.6}$$

$$\Box' S_{\mu\nu}(x-x')A_\nu(x') = \partial'_\rho \partial'_\rho S_{\mu\nu}(x-x')A_\nu(x') \Rightarrow -\partial'_\rho S_{\mu\nu}(x-x')\partial'_\rho A_\nu(x') \tag{9.7}$$

$$\partial'_\rho S_{\mu\nu}(x-x')\partial'_\rho A_\nu(x') = \partial'_\rho\left(\partial'_\nu \bar{\Delta}\partial'_\mu \Delta^1 - \bar{\Delta}\partial'_\nu \partial'_\mu \Delta^1\right)\partial'_\rho A_\nu(x') \tag{9.8}$$

$$\partial'_\rho S_{\mu\nu}(x-x')\partial'_\rho A_\nu(x') = \begin{cases}\left(\partial'_\rho \partial'_\nu \bar{\Delta}\partial'_\mu \Delta^1 + \partial'_\nu \bar{\Delta}\partial'_\rho \partial'_\mu \Delta^1\right)\partial'_\rho A_\nu(x') \\ -\left(\partial'_\rho \bar{\Delta}\partial'_\nu \partial'_\mu \Delta^1 + \bar{\Delta}\partial'_\rho \partial'_\nu \partial'_\mu \Delta^1\right)\partial'_\rho A_\nu(x')\end{cases} \tag{9.9}$$

Rearrange terms and relabel dummy variables to obtain,

$$\partial'_\rho S_{\mu\nu}(x-x')\partial'_\rho A_\nu(x') = \left(\partial'_\rho \partial'_\nu \bar{\Delta}\partial'_\mu \Delta^1 - \bar{\Delta}\partial'_\rho \partial'_\nu \partial'_\mu \Delta^1\right)\partial'_\rho A_\nu(x') - \left(\partial'_\rho \bar{\Delta}\partial'_\nu \partial'_\mu \Delta^1\right)F_{\rho\nu}(x') \tag{9.10}$$

Next perform the following integrations by parts,

$$\partial'_\rho \partial'_\nu \bar{\Delta}\partial'_\mu \Delta^1 \partial'_\rho A_\nu(x') \Rightarrow -\partial'_\nu \partial'_\rho \partial'_\rho \bar{\Delta}\partial'_\mu \Delta^1 A_\nu(x') - \partial'_\nu \partial'_\rho \bar{\Delta}\partial'_\rho \partial'_\mu \Delta^1 A_\nu(x') \tag{9.11}$$

$$\partial'_\rho \partial'_\nu \bar{\Delta}\partial'_\mu \Delta^1 \partial'_\rho A_\nu(x') \Rightarrow -\partial'_\nu \partial'_\rho \partial'_\rho \bar{\Delta}\partial'_\mu \Delta^1 A_\nu(x') + \begin{pmatrix}\partial'_\nu \bar{\Delta}\partial'_\mu \partial'_\rho \partial'_\rho \Delta^1 A_\nu(x') \\ +\partial'_\nu \bar{\Delta}\partial'_\rho \partial'_\mu \Delta^1 \partial'_\rho A_\nu(x')\end{pmatrix} \tag{9.12}$$

Use (3.23) in the above to obtain the following sequence of integration by parts,

$$\partial'_\rho \partial'_\nu \bar{\Delta}\partial'_\mu \Delta^1 \partial'_\rho A_\nu(x') \Rightarrow -\partial'_\nu\left(\partial'_\rho \partial'_\rho - m^2\right)\bar{\Delta}\partial'_\mu \Delta^1 A_\nu(x') + \partial'_\nu \bar{\Delta}\partial'_\rho \partial'_\mu \Delta^1 \partial'_\rho A_\nu(x') \tag{9.13}$$

$$\bar{\Delta}\partial'_\rho \partial'_\nu \partial'_\mu \Delta^1 \partial'_\rho A_\nu(x') \Rightarrow -\partial'_\rho \bar{\Delta}\partial'_\rho \partial'_\nu \partial'_\mu \Delta^1 A_\nu(x') - \bar{\Delta}\partial'_\nu \partial'_\mu \partial'_\rho \partial'_\rho \Delta^1 A_\nu(x') \tag{9.14}$$

$$\bar{\Delta}\partial'_\rho \partial'_\nu \partial'_\mu \Delta^1 \partial'_\rho A_\nu(x') \Rightarrow \partial'_\rho \partial'_\rho \bar{\Delta}\partial'_\nu \partial'_\mu \Delta^1 A_\nu(x') + \partial'_\rho \bar{\Delta}\partial'_\nu \partial'_\mu \Delta^1 \partial'_\rho A_\nu(x') - m^2 \bar{\Delta}\partial'_\nu \partial'_\mu \Delta^1 A_\nu(x') \tag{9.15}$$

$$\bar{\Delta}\partial'_\nu\partial'_\mu\Delta^1\partial'_\rho A_\nu(x') \Rightarrow (\partial'_\rho\partial'_\rho - m^2)\bar{\Delta}\partial'_\nu\partial'_\mu\Delta^1 A_\nu(x') + \partial'_\rho\bar{\Delta}\partial'_\nu\partial'_\mu\Delta^1\partial'_\rho A_\nu(x') \tag{9.16}$$

$$\Box' S_{\mu\nu}(x-x')A_\nu(x') \Rightarrow \partial'_\nu\left[(\Box'-m^2)\bar{\Delta}\partial'_\mu\Delta^1\right]A_\nu(x') + 2(\partial'_\rho\bar{\Delta}\partial'_\nu\partial'_\mu\Delta^1)F_{\rho\nu}(x') \tag{9.17}$$

$$\Box' S_{\mu\nu}(x-x')A_\nu(x') \Rightarrow -\left[(\Box'-m^2)\bar{\Delta}\partial'_\mu\Delta^1\right]\partial'_\nu A_\nu(x') + 2(\partial'_\rho\bar{\Delta}\partial'_\nu\partial'_\mu\Delta^1)F_{\rho\nu}(x') \tag{9.18}$$

$$2(\partial'_\rho\bar{\Delta}\partial'_\nu\partial'_\mu\Delta^1)F_{\rho\nu} \Rightarrow -2(\bar{\Delta}\partial'_\nu\partial'_\mu\Delta^1)\partial'_\rho F_{\rho\nu}(x') - 2(\bar{\Delta}\partial'_\rho\partial'_\nu\partial'_\mu\Delta^1)F_{\rho\nu}(x') \tag{9.19}$$

The last term in this last expression is zero due to the fact that $F_{\rho\nu}$ is antisymmetric. Use this in (9.5) to obtain,

$$\bar{I}_\mu(x) = 4\int\left[\left(\frac{1}{\Box}\right)\left(-\left[(\Box'-m^2)\bar{\Delta}\partial'_\mu\Delta^1\right]\partial'_\nu A_\nu(x') - 2(\bar{\Delta}\partial'_\nu\partial'_\mu\Delta^1)\partial'_\rho F_{\rho\nu}(x')\right) + (\bar{\Delta}\partial'_\nu\Delta^1)F_{\nu\mu}(x')\right]d^4x' \tag{9.20}$$

Integrate by parts to obtain,

$$\bar{I}_\mu(x) = 4\int\left[\left(-\left[(\Box'-m^2)\bar{\Delta}\partial'_\mu\Delta^1\right]\frac{\partial'_\nu}{\Box'}A_\nu(x') - 2(\bar{\Delta}\partial'_\nu\partial'_\mu\Delta^1)\left(\frac{1}{\Box'}\partial'_\rho F_{\rho\nu}(x')\right)\right) + (\bar{\Delta}\partial'_\nu\Delta^1)F_{\nu\mu}(x')\right]d^4x'$$
$$\tag{9.21}$$

which is equivalent to Eq. (4.4).

## Appendix 2.

In this appendix we will derive Eq. (6.10). Define,

$$S_{\mu\nu}(k,p) = 2k_\mu k_\nu - k_\mu p_\nu - k_\nu p_\mu - \delta_{\mu\nu}(k^2 - kp + m^2) \tag{10.1}$$

We want to examine $S_{\mu\nu}A_\nu(p)$. Use basic algebra to obtain the following two expressions,

$$2k_\mu k_\nu A_\nu = \frac{2k_\mu k_\nu p_\rho(p_\rho A_\nu - p_\nu A_\rho)}{p^2} + \frac{2k_\mu k_\nu p_\rho p_\nu A_\rho}{p^2} = \frac{2k_\mu k_\nu p_\rho(-i\tilde{F}_{\rho\nu})}{p^2} + \frac{2k_\mu(kp)(pA)}{p^2} \tag{10.2}$$

$$(-k_\nu p_\mu + kp\delta_{\mu\nu})A_\nu = k_\nu(-p_\mu A_\nu + p_\nu A_\mu) = k_\nu(-i\tilde{F}_{\nu\mu}) \tag{10.3}$$

Use these results along with (10.1) to obtain,

$$S_{\mu\nu}(k,p)A_\nu(p) = \frac{2k_\mu k_\nu p_\rho(-i\tilde{F}_{\rho\nu})}{p^2} + \frac{2k_\mu(kp)(pA)}{p^2} - k_\mu(pA) + k_\nu(-i\tilde{F}_{\nu\mu}) - \delta_{\mu\nu}(k^2 + m^2)A_\nu(p) \tag{10.4}$$

Use the following result,

$$\frac{2k_\mu(kp)(pA)}{p^2} - k_\mu(pA) = k_\mu(pA)\left(\frac{2(kp) - p^2}{p^2}\right) = \frac{k_\mu(pA)}{p^2}\left\{\begin{array}{c}-\left[(p-k)^2 + m^2\right]\\ +(k^2+m^2)\end{array}\right\} \tag{10.5}$$

in (10.4) to obtain,

$$S_{\mu\nu}(k,p)A_\nu(p) = \frac{2k_\mu k_\nu p_\rho(-i\tilde{F}_{\rho\nu})}{p^2} + \frac{k_\mu(pA)}{p^2}\left\{\begin{array}{c}-\left[(p-k)^2 + m^2\right]\\ +(k^2+m^2)\end{array}\right\} + k_\nu(-i\tilde{F}_{\nu\mu}) - \delta_{\mu\nu}(k^2+m^2)A_\nu \tag{10.6}$$

Rearrange terms to yield,

$$S_{\mu\nu}(k,p)A_\nu(p) = \left\{\begin{array}{l} \dfrac{2k_\mu k_\nu p_\rho(-i\tilde{F}_{\rho\nu})}{p^2} - \left[(p-k)^2 + m^2\right]\dfrac{k_\mu(pA)}{p^2} + k_\nu(-i\tilde{F}_{\nu\mu}) \\ + (k^2+m^2)\left[\dfrac{(k_\mu-p_\mu)(pA)+p_\mu(pA)-p^2 A_\mu}{p^2}\right] \end{array}\right\} \quad (10.7)$$

Next use $p_\mu(pA) - p^2 A_\mu = p_\nu(-i\tilde{F}_{\mu\nu}(p))$ in the above to obtain,

$$S_{\mu\nu}(k,p)A_\nu(p) = C_{\mu,GI}(k,p) + C_{\mu,NGI}(k,p) \quad (10.8)$$

where,

$$C_{\mu,GI}(k,p) = \left[\dfrac{2k_\mu k_\rho p_\nu}{p^2} + k_\nu \delta_{\rho\mu}\right](-i\tilde{F}_{\nu\rho}) + (k^2+m^2)\dfrac{p_\nu(-i\tilde{F}_{\mu\nu})}{p^2} \quad (10.9)$$

$$C_{\mu,NGI}(k,p) = \dfrac{(pA)}{p^2}\left[(k^2+m^2)(k_\mu - p_\mu) - \left[(p-k)^2+m^2\right]k_\mu\right] \quad (10.10)$$

Therefore,

$$j^{(1)}_{D\mu}(p;L) = \Pi_{\mu\nu}(p,L)A_\nu(p) = \dfrac{-e^2}{2\pi^3}\int d^4k \left\{\left[\begin{array}{c} C_{\mu,GI}(k,p) \\ +C_{\mu,NGI}(k,p) \end{array}\right]\delta(k^2+m^2)f(k-p)\theta(L-|\vec{k}|)\right\} \quad (10.11)$$

Use the relationship $x\delta(x)=0$ to obtain,

$$C_{\mu,NGI}(k,p)\delta(k^2+m^2)f(k-p) = -\dfrac{(pA)}{p^2}k_\mu\delta(k^2+m^2) \quad (10.12)$$

Therefore the potentially non-gauge invariant part is,

$$\dfrac{e^2}{2\pi^3}\dfrac{(pA)}{p^2}\int d^4k\left\{k_\mu\delta(k^2+m^2)\theta(L-|\vec{k}|)\right\} \quad (10.13)$$

Because the integrand is odd in $k$ this term equals zero. Using the above results (along with $x\delta(x)=0$) we obtain Eq. (6.10).

**Appendix 3.**

Evaluate (6.12) for the simple case where the simple case where the electrical potential has no spatial dependence. It is only dependent on time and $A_0 = 0$ and $\vec{A}(\vec{x},t) = (A_1(t),0,0)$. In momentum space $\vec{p}=0$. In this case the only non-zero term for $\tilde{F}_{\nu\rho}(p_4)$ is $\tilde{F}_{41}(p_4) = ip_4 A_1(p_4)$ and $\tilde{F}_{14}(p_4) = -ip_4 A_1(p_4)$ where $p_4 = ip_0$. Also $k_\rho p_\nu \tilde{F}_{\nu\rho} = ip_0 k_1 \tilde{F}_{41}$ and $\delta_{\rho 1}k_\nu \tilde{F}_{\nu\rho} = ik_0 \tilde{F}_{41}$ and $\delta(k^2+m^2)/\left[(k-p)^2+m^2\right] = \delta(k^2+m^2)/(-2pk+p^2)$. Use all this in (6.12) to obtain,

$$j_{DA1}(p_0;L) = \dfrac{e^2 \tilde{F}_{41}}{2\pi^3}\int d^4k\left\{\left[\dfrac{2k_1^2}{p_0} - k_0\right]\delta(k^2+m^2)P\dfrac{1}{(2p_0 k_0 - p_0^2)}\theta(L-|\vec{k}|)\right\} \quad (11.1)$$

Next use,

$$\delta(k^2+m^2) = \frac{\delta(k_0-E_k)+\delta(k_0+E_k)}{2E_k} \tag{11.2}$$

where $E_k = \sqrt{|\vec{k}|^2+m^2}$ to obtain,

$$j_{DA1}(p_0;L) = -\frac{4e^2 \tilde{F}_{41}}{2\pi^3} \int \frac{d^3k}{2E_k} \left\{ \left[\frac{E_k^2-k_1^2}{p_0}\right] P \frac{1}{(4E_k^2-p_0^2)} \right\} \theta(L-|\vec{k}|) \tag{11.3}$$

In the above using symmetry arguments we can replace $k_1^2$ with $|\vec{k}|^2/3$. In addition use $d^3k = 4\pi|\vec{k}|^2 d|\vec{k}| = 4\pi|\vec{k}|^2 (E_k/|\vec{k}|)dE_k = 4\pi E_k\sqrt{E_k^2-m^2}dE_k$, $|\vec{k}| = \sqrt{E_k^2-m^2}$ and $(E_k^2-k_1^2) = (2E_k^2+m^2)/3$ along with $\tilde{F}_{41}(p_4) = ip_4 A_1 = -p_0 A_1$ in the above to yield,

$$j_{DA1}(p_0;L) = \frac{4e^2 A_1}{3\pi^2} \int_m^\infty dE_k \left\{ P\frac{(2E_k^2+m^2)\sqrt{E_k^2-m^2}}{(4E_k^2-p_0^2)} \right\} \theta(L-|\vec{k}|) \tag{11.4}$$

Next define $z = 2E_k$ to obtain,

$$j_{DA1}(p_0;L) = \frac{e^2 A_1}{6\pi^2} \int_{2m}^\infty dz \frac{(z^2+2m^2)\sqrt{z^2-4m^2}}{(z^2-p_0^2)} \theta(L-|\vec{k}|) \tag{11.5}$$

In the above $|\vec{k}| = \sqrt{(z/2)^2-m^2}$. Next calculate $j_{DB\mu}(p_0;L)$ (Eq. (6.13)) for the assumptions given in the first paragraph of this section. We obtain,

$$j_{DB1}(x;L) = \frac{e^2 \tilde{F}_{41}}{2\pi^3} \int d^4k \left\{ \left[\frac{2k_1^2}{p_0}-k_0\right] \delta(k^2+m^2) \binom{-i\pi\varepsilon(k_0-p_0)}{\times\delta(2p_0 k_0-p_0^2)} \theta(L-|\vec{k}|) \right\} \tag{11.6}$$

Next use (11.2) in the above to obtain,

$$j_{DB1}(p_0;L) = \frac{e^2 \tilde{F}_{41}}{2\pi^3} \int \theta(L-|\vec{k}|)\frac{d^3k}{2E_k} \left\{ \begin{array}{l} \left[\frac{2k_1^2}{p_0}-E_k\right]\binom{-i\pi\varepsilon(E_k-p_0)}{\times\delta(2p_0 E_k-p_0^2)} \\ +\left[\frac{2k_1^2}{p_0}+E_k\right]\binom{-i\pi\varepsilon(-E_k-p_0)}{\times\delta(2p_0 E_k+p_0^2)} \end{array} \right\} \tag{11.7}$$

Next substitute $k_1^2 \to |\vec{k}|^2/3$ along with $|\vec{k}|^2 = E_k^2 - m^2$ and $d^3k = 4\pi E_k\sqrt{E_k^2-m^2}dE_k$ in the above to obtain,

$$j_{DB1}(p_0;L) = \frac{e^2 4\pi \tilde{F}_{41}}{2\pi^3} \int_m^\infty \frac{\sqrt{E_k^2 - m^2} dE_k}{2} \left\{ \begin{bmatrix} \frac{2(E_k^2 - m^2)}{3p_0} - E_k \end{bmatrix} \begin{pmatrix} -i\pi\varepsilon(E_k - p_0) \\ \times \delta(2p_0 E_k - p_0^2) \end{pmatrix} \\ + \begin{bmatrix} \frac{2(E_k^2 - m^2)}{3p_0} + E_k \end{bmatrix} \begin{pmatrix} i\pi\varepsilon(E_k + p_0) \\ \times \delta(2p_0 E_k + p_0^2) \end{pmatrix} \right\} \quad (11.8)$$

Next use $\delta(2p_0 E_k - p_0^2) = \delta(E_k - p_0/2)/|2p_0|$ and $\delta(2p_0 E_k + p_0^2) = \delta(E_k + p_0/2)/|2p_0|$ to obtain,

$$j_{DB1}(p_0;L) = \frac{e^2 4\pi \tilde{F}_{41}}{4\pi^3 |p_0|} \frac{\sqrt{(p_0/2)^2 - m^2} dE_k}{2} \left\{ \begin{bmatrix} \frac{2((p_0/2)^2 - m^2)}{3p_0} - \frac{p_0}{2} \end{bmatrix} \begin{pmatrix} -i\pi\varepsilon((p_0/2) - p_0) \\ \times \theta(p_0 - 2m) \end{pmatrix} \\ + \begin{bmatrix} \frac{2((p_0/2)^2 - m^2)}{3p_0} - \frac{p_0}{2} \end{bmatrix} \begin{pmatrix} -i\pi\varepsilon((p_0/2) - p_0) \\ \times \theta(-p_0 - 2m) \end{pmatrix} \right\} \quad (11.9)$$

Recall the definition of $\varepsilon$ (Eq. (3.20)) and rearrange terms to obtain,

$$j_{DB1}(p_0;L) = \frac{-i\pi e^2 |p_0| \tilde{F}_{41}}{12\pi^2} \sqrt{1 - \frac{4m^2}{p_0^2}} \left(1 + \frac{2m^2}{p_0^2}\right) \theta(|p_0| - 2m) \quad (11.10)$$

## Appendix 4.

In this appendix we derive Eq. (8.4). Define,

$$\bar{N}_\mu(x;L) = \int \bar{L}_{uv}(x-x') A_v(x') d^4 x'; \quad N_\mu(x;L) = \int L_{uv}(x-x') A_v(x') d^4 x' \quad (12.1)$$

Therefore $\hat{j}_{S\mu}^{(1)}(x) = e^2 (N_u(x-x') - \bar{N}_u(x-x'))$. First I will examine $\bar{N}_{uv}(x-x')$. Use (3.23) to obtain,

$$\delta_{\mu v} \delta^4(x-x') \Delta_L^1(x-x') A_v(x') = -\left[(\square' - m^2) \bar{\Delta}(x-x')\right] \Delta_L^1(x-x') A_\mu(x') \quad (12.2)$$

In the following we will write $\Delta_L^1$ for $\Delta_L^1(x-x')$ and similarly for $\bar{\Delta}(x-x')$ and $\Delta(x-x')$. Next integrate by parts to obtain,

$$-\left[(\square' - m^2) \bar{\Delta}\right] \Delta_L^1 A_\mu(x') \Rightarrow m^2 \bar{\Delta} \Delta_L^1 A_\mu(x') + \partial_v' \bar{\Delta} \partial_v' \Delta_L^1 A_\mu(x') + \Delta_L^1 \partial_v' \bar{\Delta} \partial_v' A_\mu(x') \quad (12.3)$$

Using integration by parts we have,

$$\bar{\Delta}(\partial_\mu' \partial_v' \Delta_L^1) A_v(x') \Rightarrow -(\partial_\mu' \bar{\Delta})(\partial_v' \Delta_L^1) A_v(x') - \bar{\Delta}(\partial_v' \Delta_L^1) \partial_\mu' A_v(x') \quad (12.4)$$

$$(\partial_\mu' \partial_v' \bar{\Delta}) \Delta_L^1 A_v(x') \Rightarrow -(\partial_v' \bar{\Delta})(\partial_\mu' \Delta_L^1) A_v(x') - (\partial_v' \bar{\Delta}) \Delta_L^1 \partial_\mu' A_v(x') \quad (12.5)$$

Use this in (8.2) to obtain,

$$\bar{L}_{\mu v}(x-x') A_v(x') \Rightarrow \left\{ \begin{matrix} \left[-2\partial_\mu' \bar{\Delta} \partial_v' \Delta_L^1 - 2\partial_v' \bar{\Delta} \partial_\mu' \Delta_L^1\right] A_v(x') + m^2 \bar{\Delta} \Delta_L^1 A_\mu(x') \\ + \partial_v' \bar{\Delta} \partial_v' \Delta_L^1 A_\mu(x') + \Delta_L^1 \partial_v' \bar{\Delta} \partial_v' A_\mu(x') - \partial_v' (\bar{\Delta} \Delta_L^1) \partial_\mu' A_v(x') \end{matrix} \right\} \quad (12.6)$$

Rearrange terms to obtain,

$$\bar{L}_{\mu\nu}(x-x')A_\nu(x') \Rightarrow \begin{cases} \left[-2\partial'_\mu \bar{\Delta}\partial'_\nu \Delta_L^1 - 2\partial'_\nu \bar{\Delta}\partial'_\mu \Delta_L^1\right]A_\nu(x') + \left(\partial'_\nu \bar{\Delta}\partial'_\nu \Delta_L^1 + m^2 \bar{\Delta}\Delta_L^1\right)A_\mu(x') \\ -\left(\partial'_\nu \Delta_L^1\right)\bar{\Delta}\partial'_\nu A_\mu(x') + \left(\bar{\Delta}\Delta_L^1\right)\partial'_\nu F_{\mu\nu}(x') \end{cases} \quad (12.7)$$

Next use,

$$\left(\partial'_\nu \bar{\Delta}\partial'_\nu \Delta_L^1 + m^2 \bar{\Delta}\Delta_L^1\right)A_\mu(x') \Rightarrow -\bar{\Delta}\partial'_\nu \Delta_L^1 \partial'_\nu A_\mu(x') \quad (12.8)$$

to obtain,

$$\bar{L}_{\mu\nu}(x-x')A_\nu(x') \Rightarrow \begin{cases} \left[-2\partial'_\mu \bar{\Delta}\partial'_\nu \Delta_L^1 - 2\partial'_\nu \bar{\Delta}\partial'_\mu \Delta_L^1\right]A_\nu(x') \\ +2\left(\partial'_\rho \bar{\Delta}\partial'_\rho \Delta_L^1 + m^2 \bar{\Delta}\Delta_L^1\right)\delta_{\mu\nu}A_\nu(x') + \left(\bar{\Delta}\Delta_L^1\right)\partial'_\nu F_{\mu\nu}(x') \end{cases} \quad (12.9)$$

Refer to Eq. (3.21)

$$\bar{L}_{\mu\nu}(x-x')A_\nu(x') \Rightarrow -\frac{1}{2}\bar{K}_{\mu\nu}(x-x')A_\nu(x') + \left(\bar{\Delta}(x-x')\Delta_L^1(x-x')\right)\partial_\nu F_{\mu\nu}(x') \quad (12.10)$$

$$\bar{N}_\mu(x;L) = \int \left(-\frac{1}{2}\bar{K}_{\mu\nu}(x-x')A_\nu(x') + \left(\bar{\Delta}(x-x')\Delta_L^1(x-x')\right)\partial_\nu F_{\mu\nu}(x')\right)d^4x' \quad (12.11)$$

Also it can be shown that,

$$N_\mu(x;L) = \int \left(-\frac{1}{2}K_{\mu\nu}(x-x')A_\nu(x') + \frac{1}{2}\left(\Delta(x-x')\Delta_L^1(x-x')\right)\partial_\nu F_{\mu\nu}(x')\right)d^4x' \quad (12.12)$$

Therefore,

$$\hat{j}_{S\mu}^{(1)}(x) = e^2 \int \left(-\frac{1}{2}\begin{pmatrix} K_{\mu\nu}(x-x') \\ -\bar{K}_{\mu\nu}(x-x') \end{pmatrix}A_\nu(x') + \begin{pmatrix} \frac{1}{2}\Delta(x-x') \\ -\bar{\Delta}(x-x') \end{pmatrix}\Delta_L^1(x-x')\partial_\nu F_{\mu\nu}(x')\right)d^4x' \quad (12.13)$$

## Appendix 5.

We want to evaluate $S_{\mu 1}(x;L)$ and $S_{\mu 2}(x;L)$ which are given by Eqs. (8.6) and (8.7), respectively. In momentum space we have,

$$S_{\mu 1}(p;L) = f_1(p)J_\mu(p) \; ; \; S_{\mu 2}(p;L) = f_2(p)J_\mu(p) \quad (13.1)$$

where,

$$f_1(p) = \frac{(2\pi)^4}{2}\int \Delta(p-k)\Delta_L^1(k)d^4k; \quad f_2(p) = (2\pi)^4 \int \bar{\Delta}(p-k)\Delta_L^1(k)d^4k; \quad (13.2)$$

and where (see Ref. [11]),

$$\bar{\Delta}(k) = \left(\frac{1}{2\pi}\right)^4 P\frac{1}{k^2+m^2}; \quad \Delta(k) = -\frac{i}{(2\pi)^3}\varepsilon(k_0)\delta(k^2+m^2); \quad \Delta^1(k) = \left(\frac{1}{2\pi}\right)^3 \delta(k^2+m^2) \quad (13.3)$$

Also, in momentum space, $\Delta_L^1(k) = \Delta^1(k)\theta(L-|\vec{k}|)$.

Solve for the case where $2m > |p_0| > 0$ and $\vec{p} = 0$. Use the above relationships to obtain,

$$f_2(p) = \frac{1}{(2\pi)^3} \int \frac{\delta(k^2+m^2)\theta(L-|\vec{k}|)}{(2p_0 k_0 - p_0^2)} d^4k; \tag{13.4}$$

Next use (11.2) to obtain,

$$f_2(p) = \frac{1}{(2\pi)^3} \int \frac{d^3k}{2E_k} \left[ \frac{1}{(2p_0 E_k - p_0^2)} - \frac{1}{(2p_0 E_k + p_0^2)} \right] \theta(L-|\vec{k}|) = \frac{-2}{(2\pi)^3} \int \frac{d^3k}{2E_k} \left[ \frac{1}{4E_k^2 - p_0^2} \right] \theta(L-|\vec{k}|) \tag{13.5}$$

Use $d^3k = 4\pi E_k \sqrt{E_k^2 - m^2}$ to obtain,

$$f_2(p) = -\frac{4\pi}{(2\pi)^3} \int_m^\infty dE_k \left[ \frac{\sqrt{E_k^2 - m^2}}{4E_k^2 - p_0^2} \right] \theta(L-|\vec{k}|) \tag{13.6}$$

where $|\vec{k}| = \sqrt{E_k^2 - m^2}$. Use $z = 2E_k$ to obtain,

$$f_2(p) = -\frac{1}{8\pi^2} \int_{2m}^\infty dz \left[ \frac{\sqrt{z^2 - 4m^2}}{z^2 - p_0^2} \right] \theta(L-|\vec{k}|) \tag{13.7}$$

This becomes,

$$f_2(p) = f_{2a}(p) + f_{2b}(p) \tag{13.8}$$

so that $S_{\mu 2}(p;L) = (f_{2a}(p) + f_{2b}(p)) J_\mu(p)$ where,

$$f_{2a}(p) = -\frac{p_0^2}{8\pi^2} \int_{2m}^\infty dz \left[ \frac{\sqrt{z^2 - 4m^2}}{z^2(z^2 - p_0^2)} \right] \theta(L-|\vec{k}|) \;;\; f_{2b}(p) = -\frac{1}{8\pi^2} \int_{2m}^\infty dz \left[ \frac{\sqrt{z^2 - 4m^2}}{z^2} \right] \theta(L-|\vec{k}|) \tag{13.9}$$

Next calculate $f_1(p)$. Use (13.3) to obtain,

$$f_1(p_0) = \frac{-i}{8\pi} \int d^4k \, \varepsilon(p_0 - k_0) \delta(2p_0 k_0 - p_0^2) \delta(k^2 + m^2) \theta(L-|\vec{k}|) \tag{13.10}$$

where I have used $\delta((p-k)^2 + m^2)\delta(k^2 + m^2) = \delta(-2pk + p^2)\delta(k^2 + m^2)$ and set $\vec{p} = 0$. Use (11.2) to obtain,

$$f_1(p_0) = \frac{-i}{8\pi} \int \frac{d^3k}{2E_k} \left[ \begin{array}{c} \varepsilon(p_0 - E_k) \delta(2p_0 E_k - p_0^2) \\ +\varepsilon(p_0 + E_k) \delta(2p_0 E_k + p_0^2) \end{array} \right] \theta(L-|\vec{k}|) \tag{13.11}$$

Use $d^3k = 4\pi E_k \sqrt{E_k^2 - m^2}$ and $\delta(2p_0 E_k - p_0^2) = \delta(E_k - (p_0/2))/|2p_0|$ and $\delta(2p_0 E_k + p_0^2) = \delta(E_k + (p_0/2))/|2p_0|$ to obtain,

$$f_1(p_0) = \frac{-i}{8\pi} \int_m^\infty dE_k \frac{\sqrt{E_k^2 - m^2}}{|p_0|} \left[ \begin{array}{c} \varepsilon(p_0 - E_k) \delta(E_k - p_0/2) \\ +\varepsilon(p_0 + E_k) \delta(E_k + p_0/2) \end{array} \right] \theta(L-|\vec{k}|) \tag{13.12}$$

This yields,

$$f_1(p_0) = \frac{-ip_0}{16\pi} \sqrt{1 - \frac{4m^2}{p_0^2}} \theta(|p_0| - 2m) \tag{13.13}$$

Recall that $S_{\mu 1}(p;L) = f_1(p) J_\mu(p)$.

## References.